\documentclass[11pt]{article}

\usepackage{graphicx} 
\usepackage{amsmath}
\usepackage{amsfonts}
\usepackage{amssymb}

\newcommand{\roughly}[1]%
    {{\mathrel{\raise.3ex\hbox{$#1$\kern-.75em\lower1ex\hbox{$\sim$}}}}}
\newcommand{\lsim}{\mathrel{\roughly<}}
\newcommand{\gsim}{\mathrel{\roughly>}}

\begin{document}

\title{\bf Rapid roll Inflation with Conformal Coupling}
\author{ Lev Kofman$^{\rm a}$ and Shinji Mukohyama$^{\rm b}$\\
\\
$^{\rm a}${\it Canadian Institute for Theoretical Astrophysics}\\
{\it University of Toronto, 60 St. George st., Toronto, ON M5S 3H8, Canada}
\\
$^{\rm b}${\it Department of Physics and 
Research Center for the Early Universe}\\
{\it The University of Tokyo, Tokyo 113-0033, Japan}}

\date{\today}

\maketitle

\begin{abstract}
 Usual inflation is realized with a slow rolling scalar field minimally
 coupled to gravity. In contrast, we consider dynamics of a scalar with
 a flat effective potential, conformally coupled to gravity.  
 Surprisingly, it contains an attractor inflationary solution with the
 rapidly rolling inflaton field. We discuss models with the conformal
 inflaton with a flat potential (including hybrid inflation). There is
 no generation of cosmological fluctuations from the conformally coupled
 inflaton. We consider realizations of modulated (inhomogeneous
 reheating) or curvaton cosmological fluctuations in these models. We
 also implement these unusual features for the popular string-theoretic
 warped inflationary scenario, based on the interacting $D3$-$\bar{D}3$
 branes. The original warped brane inflation suffers a large inflaton
 mass due to conformal coupling to $4$-dimensional gravity. Instead of 
 considering this as a problem and trying to cure it with extra
 engineering, we show that warped inflation with the conformally
 coupled, rapidly rolling inflaton is yet possible with $N=37$
 efoldings, which requires low energy scales $1$-$100$ TeV of
 inflation. Coincidentally, the same warping numerology can be
 responsible for the hierarchy. It is shown that the scalars associated
 with angular isometries of the warped geometry of compact manifold
 (e.g. $S^3$ of KS geometry) have solutions identical to conformally
 coupled modes and also cannot be responsible for cosmological
 fluctuations. We discuss other possibilities.  
\begin{flushright}
  UTAP-583, RESCEU-82/07
\end{flushright}

\end{abstract}

%\narrowtext

%======================================%
% Introduction
%======================================%
\section{Introduction} \label{sec:intro}

Realization of inflation is typically associated with a slow rolling
scalar field minimally coupled to gravity \cite{Linde:2005ht}. Is it
possible to have inflation with a non-minimally coupled scalar field,
in particular with a conformally coupled field? This question is
especially interesting in the context of superstring theory, where many
scalars are conformally coupled to gravity; or in supergravity theory,
where scalars typically have the mass of order of
$H^2$~\cite{Dine:1995uk}.

Indeed, among a number of inflationary scenarios, simple scenarios which 
come in the package with high energy physics draw an especial
attention. Recently there is significant interest to package up
inflation with the string theory. There are different models of string 
theory inflation (see, e.g. \cite{Kallosh:2007ig}). Often string
inflation models require significant fine-tuning. The most typical
problem however is the $\eta$-problem, manifested in heavy inflaton mass 
preventing slow roll of inflaton. For realization of slow-roll inflation
the effective inflaton mass should be much smaller than the Hubble
parameter during inflation, $m^2\ll H^2$. An example of this problem is
warped brane inflation (based on the configurations of interacting
branes in warped geometry) where the inflaton is conformally coupled to
the four-dimensional gravity~\cite{Kachru:2003sx}~\footnote{
Here we make a comment on the curvature coupling for the warped
brane-antibrane inflation. The face value of the curvature coupling
obtained by KKLMMT~\cite{Kachru:2003sx} is conformal, i.e. $\xi=1/6$. By 
considering inflaton-dependence of the superpotential, it is possible to 
alter  $\xi$ from $1/6$ as pointed out in the  KKLMMT and subsequent 
papers. However, we would like to stress here that curvature coupling
before including this extra ingredient is conformal and we shall
consider this value throughout this paper.}.
A scalar field $\phi$ conformally coupled to gravity acquires effective
mass term $\frac{1}{2}\xi R\phi^2$ with $\xi=1/6$. This gives the
inflaton the effective mass $m^2\simeq 12\xi H^2=2H^2$ which does not
result in the slow-roll. This problem has been considered as a stumbling
block on the way to build up successful warped brane-antibrane
inflation.

In this paper we demonstrate that in fact conformal coupling is not a
problem for realization of inflation.
Inflationary regime in this case can be realized. However, it has new
features which make it very different from other models. First,
inflation with a conformally coupled inflaton with a potential
$V(\phi)$, which is almost flat over the range of $\phi$, can be
realized as the rapid-roll inflation, contrary to the customary slow
roll inflation. To the best of our knowledge, this is the first model of
inflation with a conformally coupled inflaton. Our model can be regarded
also as a new model of the fast rolling inflaton. A model of fast roll
inflaton based on a minimally coupled scalar field was suggested in
\cite{Linde:2001ae}.

With conformal coupling, the only feature of the warped brane-antibrane
inflation needed for conformal inflation is the form of its potential:
very shallow for the most part of the inflaton rolling and changing
sharply at the end point of inflation. Therefore, application of the
conformal rapid-roll inflation is broader than the warped
brane-antibrane inflation. It includes, for instance, field-theoretic
hybrid models (with conformal inflaton) where the potential is almost
flat during inflation.

Second, the string-theory realization of rapid-roll conformal inflation
is a low energy inflation (close to the least possible energy of
inflation at $1-100$ TeV scale). As a result it requires significantly 
less efoldings $N$ than the figure $62$ typical for the GUT scale
chaotic inflation. In the context of the warped geometry, as we will
show, $N$ and the scale of inflation are directly related to the warp
factor of the throat geometry of the internal manifold. Noticeably,
lower bound for efolds of inflation $N$ coincides with the warping
needed for the Randall-Sundrum solution of the hierarchy problem.

From the viewpoint of observational cosmology, the possibility to
motivate inflation which lasts just to provide homogeneity/isotropy only
within the observable horizon (but not beyond it) is interesting with
respect to the low multi-poles anomalies found in the COBE and WMAP CMB
temperature anisotropies. There are attempts to build up the models with
the spectra of primordial fluctuations truncated at the horizon scales
(for instance see \cite{Contaldi:2003zv}), but not much motivation for
the choice of the scale of truncation was found. The model of conformal
inflation naturally gives truncation of fluctuations at the horizon
scales.

The third feature of our model is different origin of its cosmological
fluctuations. A conformally coupled inflaton cannot be responsible for
generation of primordial fluctuations. As we shall show, fluctuations
also cannot be generated from the angular degrees of freedom, associated
with the angular positions of branes in the bulk Klebanov-Strassler
throat geometry. We propose modulated (inhomogeneous reheating)
fluctuations~\cite{Dvali:2003em,Kofman:2003nx} or curvaton
fluctuations~\cite{Enqvist:2001zp,Lyth:2001nq} of the scalars of the SM
sector or other moduli fields.

In the rest of the paper we will address these features of the model.

%======================================%
% Inflation 
%======================================%
\section{Inflation from Conformally-Coupled Scalar with Flat Potential}

Conventional inflationary models are based on minimally coupled scalar
fields. Indeed, slow roll inflation requires relatively small effective
inflaton mass  $m^2\ll H^2$, while conformal coupling violates this
condition.

Let us nonetheless consider a conformally coupled scalar field with a
very simple flat potential. This is relevant for a couple of interesting
inflationary models. In the warped brane inflation the inflaton
candidate is an open string modulus, brane-antibrane distance. Denote
this scalar field in the four-dimensional effective theory as
$\phi$. Brane-antibrane interaction results  in the four-dimensional
scalar field potential~\cite{Kachru:2003sx}
%============< EQUATION >==============%
%
\begin{equation}\label{potential}
 V = V_0\left[ 1 - \left(\frac{M_p\Delta}{\phi}\right)^4\right] \ ,
\end{equation}
%======================================%
where $M_p=1/\sqrt{8\pi G_N}$ ($\simeq 2\times 10^{18}GeV$) is the
reduced Planck mass, $\Delta$ is a dimensionless constant, $V_0$ is
related to the warp factor $e^{-A}$ as $V_0\sim T_3 e^{-4A}$ and $T_3$
is the $D3$-brane tension. Except for small values of $\phi$, this
potential is very flat. Another example of almost flat potential is the
usual field-theoretic hybrid inflation.

As the first step, consider dynamics of the FRW universe dominated by a 
conformally-coupled scalar field with a potential which can be
approximated by a constant. Although the potential is flat, scalar field 
dynamics is significant due to the conformal coupling to the curvature,
which can be interpreted as induced mass of the scalar
field. Non-trivial observation is that de Sitter spacetime is
self-consisted solution of this system. Moreover, inflation is an
attractor solution  for a FRW universe with a conformally-coupled scalar
field with $V=const$.

For a scalar field $\phi$ described by the action
%============< EQUATION >==============%
%
\begin{equation}\label{lagrangian}
 I = \int d^4x\sqrt{-g}
  \left[ \frac{M_p^2R}{2}
   -\frac{1}{2}\partial^{\mu}\phi\partial_{\mu}\phi
   -V(\phi) - \frac{\xi}{2}R\phi^2\right] \ ,
\end{equation}
%======================================%
where $\xi$ is a coupling to gravity, the equations of motion are
%============< EQUATION >==============%
%
\begin{equation}
 \left(M_p^2-\xi\phi^2\right)G_{\mu\nu}
  = \partial_{\mu}\phi\partial_{\nu}\phi
  -\frac{1}{2}\partial^{\rho}\phi\partial_{\rho}\phi g_{\mu\nu}
  -\xi\left[\nabla_{\mu}\nabla_{\nu}(\phi^2)
       -\nabla^{\rho}\nabla_{\rho}(\phi^2)g_{\mu\nu}\right]
  - V(\phi)g_{\mu\nu} \ ,
\end{equation}
%======================================%
and
%============< EQUATION >==============%
%
\begin{equation}
 \nabla^{\mu}\nabla_{\mu}\phi - V'(\phi) - \xi R\phi = 0 \ .
\end{equation}
%======================================%

In the FRW background
%============< EQUATION >==============%
%
\begin{equation}
 ds^2 = -dt^2 + a(t)^2d{\bf x}^2
\end{equation}
%======================================%
these equations are reduced to
%============< EQUATION >==============%
%
\begin{equation}\label{eq1}
 3\left(M_p^2-\xi\phi^2\right)H^2 = 
  \frac{1}{2}\dot{\phi}^2 + 6\xi H \phi\dot{\phi} + V(\phi),
\end{equation}
%======================================%
and
%============< EQUATION >==============%
%
\begin{equation}\label{eq2}
 \ddot{\phi} + 3H\dot{\phi} + 6\xi(\dot{H}+2H^2)\phi + V'(\phi) = 0.
\end{equation}
%======================================%

With $\xi=1/6$ and $V(\phi)=V_0$, it is convenient to
introduce the new variable $\varphi$ by
%============< EQUATION >==============%
%
\begin{equation}
 \phi = \frac{\varphi}{a}.
\end{equation}
%======================================%
With this variable, the equations of motion are as simple as
%============< EQUATION >==============%
%
\begin{equation}
 {a'}^2 = H_0^2\left(a^4+\frac{{\varphi'}^2}{2V_0}\right), \quad
 \varphi'' = 0,
\end{equation}
%======================================%
where 
%============< EQUATION >==============%
%
\begin{equation}
 H_0 \equiv \sqrt{\frac{V_0}{3M_p^2}}, 
\end{equation}
%======================================%
and a prime denotes the derivative w.r.t. the conformal time $\eta$ 
defined by $dt=ad\eta$. The second equation implies that 
%============< EQUATION >==============%
%
\begin{equation}
 \varphi = \varphi_* + v\eta,
\end{equation}
%======================================%
where $\varphi_*$ and $v$ are constants. Thus, the first equation
becomes 
%============< EQUATION >==============%
%
\begin{equation}\label{prime}
 {a'}^2 = H_0^2\left(a^4+a_0^4\right), 
\end{equation}
%======================================%
where $a_0^4\equiv\frac{v^2}{2V_0}$. For the expanding branch ($a'>0$),
solution of Eq.~(\ref{prime}) is 
%============< EQUATION >==============%
%
\begin{equation}
 H_0(\eta_0-\eta) = 
  \int_a^{\infty}\frac{dx}{\sqrt{x^4+a_0^4}} \equiv F(a) \ ,
 \label{sol}
\end{equation}
%======================================%
where $\eta_0=const$.

For $a\gg a_0$, the function $F(a)$ is well approximated as 
$ F(a)\simeq\frac{1}{a}$. Thus, from Eq.~(\ref{sol}) we immediately
obtain 
%============< EQUATION >==============%
%
\begin{equation}\label{scale}
 a \simeq \frac{1}{H_0(\eta_0-\eta)}, \quad 
 \varphi \simeq \varphi_0 \quad
 \mbox{for }\eta\to \eta_0-0,
\end{equation}
%======================================%
where $\varphi_0\equiv\varphi_*+v\eta_0$. The scale factor (\ref{scale})
corresponds to the de Sitter solution. Therefore, we have shown that
inflation  with the Hubble expansion rate $H_0$ is an attractor solution
of Eqs.~(\ref{eq1}), (\ref{eq2}) with $\xi=1/6$. In
Sec.~\ref{sec:portrait}, we shall illustrate this attractor behavior for
the potential (\ref{potential}) by using the phase portrait method.

%======================================%
% Conditions for inflation
%======================================%
\section{Condition for the Rapid Roll  Inflation}

As we have shown above, a conformally coupled scalar field can lead to 
inflation. During inflation, however, the inflaton evolves as 
%============< EQUATION >==============%
%
\begin{equation}\label{fast}
 \phi \approx  \frac{\varphi_0}{a} ,
\end{equation}
%======================================%
with the velocity
%============< EQUATION >==============%
%
\begin{equation}\label{speed}
\dot  \phi \approx -H \phi \ .
\end{equation}
%======================================%
This is rapid roll inflation. The number of efolds  $N=\int dt H$ for
this rapid roll inflation is reduced to 
%============< EQUATION >==============%
%
\begin{equation}\label{efolds}
N=-\int \frac{d \phi}{\phi}=\ln\frac{\phi_i}{\phi_f} \ ,
\end{equation}
%======================================%
where $\phi_i$ and $\phi_f$ are the values of the inflaton at the
beginning and end of inflation. As we will show in the next section, low
energy inflation can be realized for the warped geometry brane
inflation.

In the previous section we have shown how conformal inflation can be
realized  with the flat potential. In this section we derive the
conditions for a general potential $V(\phi)$ to realize conformal
inflation.

For a minimally coupled inflaton, conditions for the potential $V(\phi)$
to provide inflation can be formulated in terms of smallness of the
slow-roll parameters 
%============< EQUATION >==============%
%
\begin{equation}\label{small}
 \epsilon \equiv \frac{M_p^2}{2}\left(\frac{V'}{V}\right)^2
  \ , \,\, \,  \eta \equiv \frac{M_p^2V''}{V}.
\end{equation}
%======================================%

We will discuss their generalization for the rapid roll inflaton. For an
arbitrary  potential $V(\phi)$, the equation of motion and the Friedman
equation for the conformal inflaton can be re-written as 
%============< EQUATION >==============%
%
\begin{equation}
 H^2 = \frac{1}{3M_p^2}\left(\frac{1}{2}\pi^2+V\right), \quad 
 \dot{\pi} + 2H\pi + V' = 0,  \label{eqn:eom-pi}
\end{equation}
%======================================%
where
%============< EQUATION >==============%
%
\begin{equation}
 \pi \equiv \dot{\phi} + H\phi. 
\end{equation}
%======================================%

Inflation corresponds to the slow variation of the Hubble parameter
$H$. Therefore it is instructive to construct 
%============< EQUATION >==============%
%
\begin{equation}
 -\frac{\dot{H}}{H^2} = \frac{\pi^2+V'\phi/2}{\pi^2/2+V}. 
  \label{eqn:dotH-over-H2}
\end{equation}
%======================================%
It is convenient to introduce the following combinations 
%============< EQUATION >==============%
%
\begin{equation}
 \tilde{\epsilon}\equiv \frac{V'\phi}{2V}, \quad
 \eta_c\equiv \eta + \frac{c+2}{3}\left[\frac{V''\phi}{V'}+c\right],
  \label{cond}
\end{equation}
%======================================%
where $c$ is a dimensionless constant chosen so that $|\eta_c|$ is
minimized in a range of $\phi$.

In Appendix A we derive the conditions for the conformal inflation: 
%============< EQUATION >==============%
%
\begin{eqnarray}
 \epsilon & \ll & 1, \label{eqn:epsilon} \\
 |\tilde{\epsilon}| & \ll & 1,   \label{eqn:tilde-epsilon}\\
 |\eta_c| & \ll & 1.
  \label{eqn:tilde-eta_c}
\end{eqnarray}
%======================================%
Under these conditions the equations motion are approximated as 
%============< EQUATION >==============%
%
\begin{eqnarray}
 H^2 & \simeq & \frac{1}{3M_p^2}V, \nonumber\\
 (2+c) H\pi & \simeq & -V', 
\end{eqnarray}
%======================================%
and the Hubble expansion rate $H$ is slowly varying with time,
i.e. $|\dot{H}/H^2|\ll 1$.

%======================================%
% Phase Portrait
%======================================%
\section{Phase Portrait for Conformal Inflation}
\label{sec:portrait}

In the KKLMMT setup for the warped brane inflation~\cite{Kachru:2003sx},
an inflaton emerges as a conformal field. It has been believed that the
conformal coupling should spoil the flatness of the inflaton potential
and that additional severe fine-tuning should be necessary. In this
section we shall show that, contrary to the folklore, inflation in this
model can be realized without further fine-tuning.

The inflaton in the warped  setup is a conformally coupled scalar field
with the potential (\ref{potential}). For this potential, the condition
(\ref{eqn:tilde-eta_c}) with $c=5$ is reduced just to $|\eta|\ll 1$. 
Thus, the conditions for inflation (\ref{eqn:epsilon}),
(\ref{eqn:tilde-epsilon}) and (\ref{eqn:tilde-eta_c}) can be written as 
%============< EQUATION >==============%
%
\begin{equation}
 \left|\frac{\phi}{M_p}\right| \gg { Max} [ \Delta, \Delta^{2/3}].
  \label{eqn:inflationary-regime}
\end{equation}
%======================================%

In this section we will analyze generic solutions of the
Eqs.~(\ref{eqn:eom-pi}) with the potential (\ref{potential}). We can
study scalar field/gravity dynamics in great details using the powerful
phase portrait method, which shows the character of {\it all} solutions 
for $a(t)$ and $\phi(t)$ as trajectories in the two-dimensional phase
space. For the minimally coupled scalar field, convenient coordinates
for the phase portrait are $(\phi,\dot{\phi})$. Phase portraits for
monomic potentials contain separatrices which attract most of the
trajectories~\cite{Belinsky:1985zd,Kofman:1985aw}. These separatrices
correspond to the inflationary regime of slow rolling.

In the case of a conformal scalar field, the qualitative behavior of
the scalar field/gravity system can be also studied  by drawing the
phase portrait. For this purpose, we introduce dimensionless variables
$q$ and $p$ and the dimensionless time coordinate $\tau$ as 
%============< EQUATION >==============%
%
\begin{equation}
 q \equiv \frac{\phi}{M_p}, \quad p \equiv \frac{\pi}{\sqrt{V_0}}, 
  \quad \tau \equiv \frac{\sqrt{V_0}}{M_p}t.
\end{equation}
%======================================%
We will deal with the two-dimensional phase portrait in terms of 
$(\phi, \pi)$, or with its dimensionless copy in terms of $(q,p)$.

\begin{figure}
 \begin{center}
  \includegraphics[trim = 0 0 0 0 ,scale=1.0, clip]{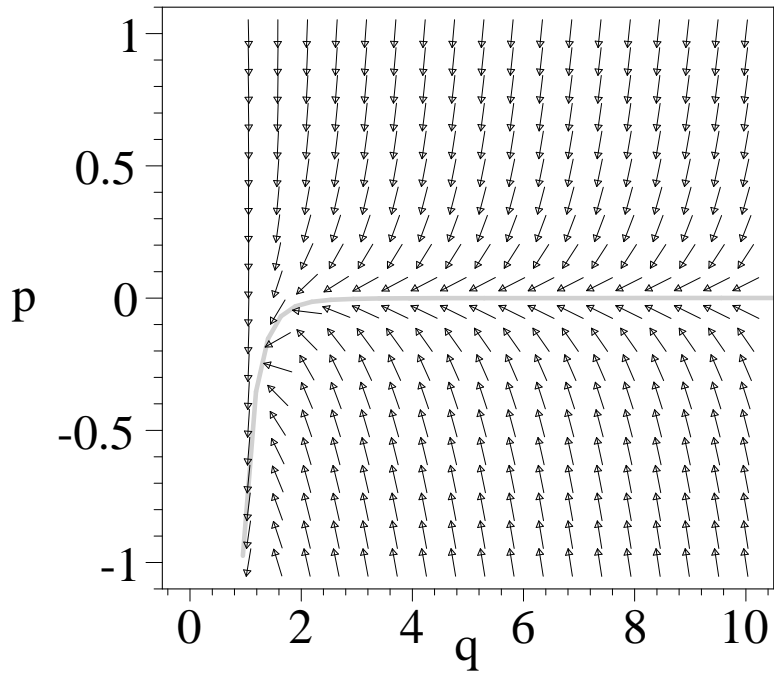}
 \end{center}
 \caption{\label{fig:portrait}
 Phase portrait for the system of equations (\ref{eqn:1st-order-eq})
 with $\Delta=1$. The horizontal axis is $q$ and the vertical axis is
 $p$. 
 }
\end{figure}

The equations of motion (\ref{eqn:eom-pi}) can be  written in terms of
$(q,p)$ as 
%============< EQUATION >==============%
%
\begin{eqnarray}
 \partial_{\tau}q & = & p - \frac{1}{\sqrt{3}}hq, \nonumber\\
 \partial_{\tau}p & = & -\frac{2}{\sqrt{3}}hp - \frac{4\Delta^4}{q^5} \ ,
  \label{eqn:1st-order-eq}
\end{eqnarray}
%======================================%
where $h\equiv \sqrt{3M_p^2/V_0}H$. The constraint equation is written
in terms of $q$, $p$ and $h$ as 
%============< EQUATION >==============%
%
\begin{equation}
 h = \sqrt{1+\frac{1}{2}p^2-\left(\frac{\Delta}{q}\right)^4}. 
  \label{eqn:constraint}
\end{equation}
%======================================%
Here, we have chosen the expanding branch $h>0$. 
Figure~\ref{fig:portrait} shows the phase portrait of the system
(\ref{eqn:1st-order-eq}) with (\ref{eqn:constraint}). There is a
separatric with $|p|\ll 1$. One can easily see that most of the
solutions quickly approaches this inflationary separatrices.

%============< FIGURE >==============%
%       conformal_portrait.eps
%======================================%

%======================================%
% Efoldings
%======================================%

\section{Efoldings of inflation  and Mass Hierarchy}

At first glance there is no common ground for the mass hierarchy and
duration of inflation. We give a surprising example where both of those
emerge from the same underlying physics.

Consider warped brane inflation. Suppose inflation begins at the initial
field value $\phi_i$, and ends at the end point of inflation $\phi_f$. 
According to Eq.~(\ref{efolds}), number of efolds of the rapid roll
inflation is $\ln({\phi_i/\phi_e})$. In the context of the
warped geometry the modulus field $\phi$ is related to the radial
coordinate of the throat $y$ as $\phi=e^{-y/R}$, where $R$ is a radius
of the AdS which well approximates the KS throat geometry. Thus the
number of efolds is 
%============< EQUATION >==============%
%
\begin{equation}\label{y}
 N=(y_i-y_e)/R 
\end{equation}
%======================================%
and is bounded from above as
%============< EQUATION >==============%
%
\begin{equation}
 N \lsim A, \label{eqn:NlsimA}
\end{equation}
%======================================%
where $A\equiv (y_{top}-y_{tip})/R$, and $y_{top}$ and $y_{tip}$ are
values of the radial coordinate at the top and the tip of the throat,
respectively. Note that $e^{-A}$ is the warp factor at the tip of the
throat geometry. The value of the warp factor $e^{-A}$ depends on the
energy scale associated with the throat. For example, the warp factor
$\sim e^{-37}$ is related to the Standard Model, TeV scale throat. (See
(\ref{hierarchy}) below with $M\simeq TeV$.) Thus, in single-throat
scenarios where the brane-antibrane pair is also in the TeV scale
throat, the inequality (\ref{eqn:NlsimA}) says that the number of
efolding $N$ is less than $37$.

We shall see below that even in multi-throat scenarios where the scale
of inflation and the electroweak scale are not necessarily
related~\footnote{For problems with multi-throat scenarios and a
possible resolution, see \cite{Mukohyama:2007ig}.}, the number of
efoldings $N_*$ corresponding to the wavelength of the present
observable horizon is bounded from above. Actually, we shall obtain the
bound $N_*\lsim 34$ applicable to both single- and multi-throat
scenarios. Moreover, we shall show that the scale of inflation
$M=V^{1/4}$ is also bounded from above by $\sim 10^6GeV$. This bound
also applies to both single- and multi-throat scenarios.

One might think that $N_*\lsim 34$ would be too small for successful 
inflation. GUT scale inflation requires at the minimum about $60$ efolds
or so. However, low energy inflation needs significantly smaller
$N_*$. Indeed, recall that number of efolds which corresponds to the
wavelength of the present-day observable horizon is related to the
energy scale of inflation $M=V^{1/4}$ as 
%============< EQUATION >==============%
%
\begin{equation}\label{eqn:efold}
 N_* \approx 62-\ln \frac{10^{16} GeV}{M}-\Delta_r \ ,
\end{equation}
%======================================%
where $\Delta_r=\frac{4}{3}\ln \frac{M}{T_{r}}$ and $T_r$ is reheat
temperature. Here, for simplicity we have assumed that the energy scale
of inflation $M=V^{1/4}$ is not significantly changing during
inflation. Depending on the (p)reheating scenario, $\Delta_r$ varies in
the range
%============< EQUATION >==============%
%
\begin{equation}
 1 \lsim \Delta_r \lsim 10.\label{eqn:Deltar}
\end{equation}
%======================================%
For example, let us take TeV as the scale of inflation, then number
of efolds in (\ref{eqn:efold}) for the TeV scale inflation can be pushed
to the lowest value $N_* \sim 30$.

In order to derive the upper bounds on $N_*$ and $M$, let us recall the 
warping (Randall-Sundrum) solution of the hierarchy problem \cite{Randall:1999ee}
between the Planck scale $M_{Pl}$ ($=\sqrt{8\pi}M_p\simeq 10^{19}GeV$)
and low energy scale $M$, based on the warped AdS geometry 
%============< EQUATION >==============%
%
\begin{equation}\label{hierarchy}
 M \simeq  M_{Pl}e^{-A}. 
\end{equation}
%======================================%
Suppose that the scale of inflation $M=V^{1/4}$ is determined in this
way. Then, by substituting this to (\ref{eqn:efold}) we obtain 
%============< EQUATION >==============%
%
\begin{equation}
 A + N_* \simeq 69 - \Delta_r. 
\end{equation}
%======================================%
Thus, (\ref{eqn:Deltar}) implies that
%============< EQUATION >==============%
%
\begin{equation}
 59 \lsim A+N_* \lsim 68. 
\end{equation}
%======================================%
Combining this with (\ref{eqn:NlsimA}) and using (\ref{hierarchy})
again, we obtain the bounds 
%============< EQUATION >==============%
%
\begin{equation}
 N_* \lsim 34, \quad M \lsim 10^6GeV. 
  \label{eqn:bounds}
\end{equation}
%======================================%
These bounds apply to both single- and multi-throat scenarios and tell 
that the conformal inflation in warped geometry is low-energy
inflation.

Let us now see that the conformal inflation is compatible with
$M\simeq TeV$. For this purpose, note that the inequality
(\ref{eqn:NlsimA}) is not saturated in general. Inflation does not have
to start exactly at the top of the throat and ends before the
tip. Moreover, the efoldings $N_*$ corresponding to the wavelength of
the present horizon be smaller than the total efoldings $N$. Thus, let
us introduce the offset $\Delta_A$ as
%============< EQUATION >==============%
%
\begin{equation}
 \Delta_A \equiv A - N_* \quad (>0). 
\end{equation}
%======================================%
Hence, we obtain
%============< EQUATION >==============%
%
\begin{equation}
 29.5 + \frac{1}{2}\Delta_A \lsim A \lsim 34 + \frac{1}{2}\Delta_A,
\end{equation}
%======================================%
where the first inequality is saturated with $\Delta_r\simeq 10$ and the
second with $\Delta_r\simeq 1$. Thus, if $6\lsim\Delta_A\lsim 15$ then 
$M\simeq TeV$ is possible with $A\simeq 37$. Note that the formula
(\ref{eqn:efold}) relates the number of efolds and the horizon-size
wavelength of cosmological fluctuation. It assumes that the universe is
at the inflationary stage at $N_*$ efolds from the end of
inflation. However, it is more comfortable to have the total number of
efolds $N$ slightly bigger, to dilute potential inhomogeneities which
could exit prior inflation. Thus, from this point of view, it is
plausible to have $\Delta_A\gsim 6$. On this other hand, 
$\Delta_A\lsim 15$ implies that the total number of efolds $N$ is not
significantly larger than the number of efolds $N_*$ corresponding to
the wavelength of the present horizon scale: $N-N_*\lsim 15$.

%======================================%
% Other examples 
%======================================%
\section{Other Examples of Conformal Inflation}

Besides the warped brane inflation of the string theory, we can give
other, phenomenological examples in which the conformal inflation is
realized. Consider the potential $V(\phi)$, which is flat for large
values of $\phi$, $V =M^4=constant$, but abruptly drops to zero at some 
value $\phi_{f}$. Let us estimate energy scale $M$ and correspondingly
the number of efolds in this model. Number of efolds $N$ is given by the
formula (\ref{efolds}). The choice of initial value $\phi_i$ is a matter
of taste: string theorist would put $\phi_i \sim M_p$ while
phenomenologically we can take it much larger. 

Let us estimate the lower bound on $\phi_f$. In order to have effective
mass $V_{\,\phi\phi}$ to be less than $M_p^2$ everywhere, we estimate
$\phi_f^2 M_p^2 \gsim M^4$. On the other hand, we have 
$H^2 M_p^2 \sim M^4$ so that $\phi_f \gsim H$. If we take 
$\phi_i \sim M_p$, then we have 
$N\lsim\ln\frac{M_p}{H}\sim 2\ln\frac{M_p}{M}$. On the other hand, the
constraint on $N$ is related to the energy scale of inflation via the 
formula (\ref{eqn:efold}). Combining both formulas for $N$, we obtain
a condition for $M$. The result is $M\lsim 10^8-10^{10}$ GeV.

Now, if we increase $\phi_i$, the value of $M$ will increase. Increase
of  $\phi_f$, in contrast, decreases  $M$. 

Similar estimation can be made for the hybrid inflationary model with
the potential 
$V(\phi,\sigma)=
\frac{\lambda}{4}(\sigma^2-v^2)^2+\frac{g^2}{2}\phi^2\sigma^2$ 
plus conformal couplings. In this case $\phi_f$ is associated with the
critical value $\phi_c=\frac{\sqrt{\lambda}}{g} v$. The result for $M$
will depends on $\phi_i$ and the ratio $g/\lambda^{1/4}$.

%======================================%
% Fluctuations
%======================================%
\section{Cosmological Fluctuations in Conformal Inflation}

At this point the model we are investigating is far from being
complete. We have not elaborated the picture of cosmological
fluctuations in the models of conformal inflation.  In this section we
just discuss some ideas about generation of cosmological fluctuations in
the model. Conformally coupling inflaton cannot be responsible for
generation of primordial fluctuations. Indeed, conformal inflaton with
the constant potential is described by the equation
$\left(\nabla_{\mu}\nabla^{\mu}+\frac{1}{6}R \right)=0$, which can be
mapped by conformal transformation $\phi =\varphi /a$ to the wave
equation in the flat space-time. Thus, conformal inflaton fluctuations
cannot be generated during inflation.

Cosmological fluctuations, however, can be generated as the modulated
(e.g. inhomogeneous reheating)
fluctuations~\cite{Dvali:2003em,Kofman:2003nx} or curvaton 
fluctuations~\cite{Enqvist:2001zp,Lyth:2001nq}. Obvious candidate for
these are fluctuations of the angular degrees of freedom. Indeed,
setting of the warped brane-antibrane inflation includes $D3$ brane and
anti $D3$ brane in six dimensional Klebanov-Strassler throat
geometry~\cite{Klebanov:2000hb}. While radial distance of
$D3$-$\bar{D}3$ pair acts as an inflaton, its five bulk angular
coordinates are other moduli fields. It turns out that angular 
fluctuations are effectively conformal scalar fields and cannot be
responsible for the modulated fluctuations. This is demonstrated in the
Appendix B. If inflaton is not conformal, angular fluctuations are also
non-conformal and generated during inflation. Similarly, modulated
fluctuations in principle can be provided by  the scalars of SM sector
(the Higgs fields) or from the moduli field coupled with the SM
sector. For this we need a light scalar field with the effective mass
smaller than the Hubble parameter. If we deal with the fields with the
mass term $m^2\phi^2$ which is not changing during the inflation and
reheating, we encounter here new difficulties. Indeed, the Hubble
parameter in this low-energy inflation is very  small, $H\sim M^2/M_p$,
$10^{-3}eV < H < 10 eV$. Potential candidates for very light scalars
would be the K\"ahler moduli or axions associated with the four-cycles
of the Calabi-Yau manifold. K\"ahler moduli $\tau$  may have very flat
potential at large values of $\tau$, while axion is very light. Another
potential candidate of the light scalar is the scalar mode that is the
super partner of the Goldstone mode associated with $U(1)_B$ symmetry
breaking in KS geometry~\cite{Gubser:2004qj,Gubser:2004tf}. It is
massless in the limit of infinite throat but shall acquire mass when the
throat is compactified. Other possibilities is to have the scalar field
potential such that moduli or curvaton scalar is very light scalars
during inflation but became massive at lower energies after inflation. 
The amplitude of modulated or curvaton fluctuations is proportional to
$H/\phi_*$, where $\phi_*$ is the mean value amplitude of the
moduli/curvaton during inflation. One has to adjust $\phi_*$ to be small
to reach required level of fluctuations $10^{-5}$. Potential modulated
or curvaton fluctuations require further investigations which are out of
the scope of this paper.

In other models of conformal inflation, where its energy scale is larger,
the value $\phi_*$ can be much larger. For instance, for $M\sim 10^{10}$
GeV, $\phi_*\sim 10^7$ GeV.

%======================================%
% Summary
%======================================%
\section{Summary}

Contrary to the folklore that conformal coupling should spoil
inflationary behavior, we have shown that a conformally coupled scalar 
field can actually drive inflation. This surprising result potentially
has significant impacts on development of the phenomenological
inflationary models as well as inflationary models in string theory
since an inflaton for warped brane inflation typically has conformal
coupling. Indeed, the inflaton potential (without further fine-tuning)
for warped brane inflation (\ref{potential}) satisfies all the
conditions for conformal inflation (\ref{eqn:epsilon}),
(\ref{eqn:tilde-epsilon}) and (\ref{eqn:tilde-eta_c}) in the regime 
(\ref{eqn:inflationary-regime}). As shown in Figure~\ref{fig:portrait},
we have confirmed the onset of inflationary behavior (for a wide range  
of initial conditions) by using the phase portrait method.

We have shown the following three features of the conformal inflation:
\begin{itemize}
\item[i)]
	The conformal inflation is a rapid  roll inflation. The inflaton
	rolls as $\phi=\varphi_0/a$, where $\varphi_0$ is a constant and
	$a$ is the scale factor. 
\item[ii)]
	For the conformal inflation in the context of warped extra
	dimensions, the number of inflationary efoldings and the scale
	of inflation are related to the warp factor of the throat
	geometry. As a result, the scale of inflation $M$ is bounded
	from above as $M\lsim 10^6GeV$. It has similar bound in the
	context of the hybrid inflation (unless $\phi$ is not bounded by
	$M_p$ from above). 
\item[iii)]
	Cosmological fluctuations originated from modulated or curvaton
	fluctuations. 
\end{itemize}
Since the conformal inflation arises rather naturally in the context of
warped brane inflation in string theory, it is certainly worthwhile 
investigating more details of its properties.

Another interesting question is about how realization of inflation
depends on the value of the curvature coupling $\xi$. The best way to
study this issue apparently would be to construct the phase portrait of
the dynamical system (\ref{eq1})-(\ref{eq2}), and to seek for the 
inflationary separatrices. For instance, as it was done in 
\cite{Felder:2002jk}. Looking at the constraint equation (\ref{eq1}), we 
notice that at least the cases $\xi=0$ and $1/6$ have nice properties:
the kinetic terms can be separated from other terms, as $\dot \phi^2$
for $\xi=0$ and as $\pi^2$ for $\xi=1/6$. We leave this issue for the
future investigation.

%======================================%
% Acknowledgements
%======================================%
\section*{Acknowledgments}

We are grateful to N.~Barnaby, G.~Felder, S.~Kachru, I.~Klebanov,
A.~Linde, K.~Maeda, D.~Pogosyan,  M.~Sasaki and J.~Yokoyama for useful
discussions. LK was supported by NSERC and CIAR. SM was supported by
MEXT through a Grant-in-Aid for Young Scientists (B) No.~17740134, and
by JSPS through a Grant-in-Aid for Creative Scientific Research
No.~19GS0219 and through a Grant-in-Aid for Scientific Research (B)
No.~19340054. LK thanks SITP and KIPAC for hospitality and support at
Stanford. SM thanks CITA and APCTP for hospitality.

%======================================%
%<<<<<<<<<<<< APPENDICES >>>>>>>>>>>>>>%
%======================================%

%======================================%
% Condition for inflation
%======================================%
\section*{Appendix A: Conditions for Inflation}

This Appendix continues discussion of the conditions for conformal
inflation. We suppose that the equations of motion (\ref{eqn:eom-pi})
are approximated by
%============< EQUATION >==============%
%
\begin{eqnarray}
 H^2 & \simeq & \frac{1}{3M_p^2}V, \nonumber\\
 \tilde{c} H\pi & \simeq & -V', 
  \label{eqn:approx-eom-pi}
\end{eqnarray}
%======================================%
where $\tilde{c}$ is a constant being determined in the following
argument.

Validity of the first approximate equation requires that $\pi^2/V$ be
sufficiently smaller than unity. Since 
%============< EQUATION >==============%
%
\begin{equation}
 \frac{\pi^2}{V} = \frac{(\tilde{c}H\pi)^2}{\tilde{c}^2H^2V}
  \simeq \frac{6}{\tilde{c}^2}\epsilon, \label{eqn:pi2-over-V} \ .
\end{equation}
%======================================%
This demands that $\epsilon\ll 1$. Before seeking the consistency
condition for the second approximate equation, let us consider the
condition for the Hubble expansion rate $H$ to be almost constant. This
condition is written as $|\dot{H}/H^2|\ll 1$, where $\dot{H}/H^2$ is
estimated by using (\ref{eqn:dotH-over-H2}) and (\ref{eqn:pi2-over-V})
as 
%============< EQUATION >==============%
%
\begin{equation}
 -\frac{\dot{H}}{H^2} = \frac{\pi^2/V+\tilde{\epsilon}}{\pi^2/2V+1} 
  \simeq
  \frac{6\epsilon/\tilde{c}^2+\tilde{\epsilon}}{1+3\epsilon/\tilde{c}^2}
  \simeq \tilde{\epsilon} \ .
\end{equation}
%======================================%
and we have used the condition $\epsilon\ll 1$. Thus, in order for the
Hubble expansion rate $H$ to be almost constant, it is required that
$|\tilde{\epsilon}|\ll 1$.

Now let us seek the consistency condition for the second approximate
equation. Since it ignores $\dot{\pi}-(\tilde{c}-2)H\pi$ relative to
$\tilde{c}H\pi$, we have to estimate
$|[\dot{\pi}-(\tilde{c}-2)H\pi]/(\tilde{c}H\pi)|$ and to demand that it
be sufficiently smaller than unity. For this purpose let us take the
time derivative of the second approximate equation as 
%============< EQUATION >==============%
%
\begin{equation}
 H\dot{\pi}+\dot{H}\pi \simeq -\frac{1}{\tilde{c}}V''(\pi-H\phi). 
\end{equation}
%======================================%
Hence, 
%============< EQUATION >==============%
%
\begin{equation}
\dot{\pi} - (\tilde{c}-2)H\pi
 \simeq 
 -\frac{V''\pi}{\tilde{c}H} - \frac{\dot{H}\pi}{H}
 + \frac{V''\phi}{\tilde{c}} - (\tilde{c}-2)H\pi. 
\end{equation}
%======================================%
Thus, by using the approximate equations and the condition
$|\dot{H}/H^2|\ll 1$ demanded above, we obtain 
%============< EQUATION >==============%
%
\begin{eqnarray}
\frac{\dot{\pi} - (\tilde{c}-2)H\pi}{\tilde{c}H\pi}
 & \simeq &
 -\frac{V''}{\tilde{c}^2H^2} - \frac{\dot{H}}{\tilde{c}H^2}
 + \frac{V''\phi}{\tilde{c}^2H\pi} - \frac{\tilde{c}-2}{\tilde{c}}
 \nonumber\\
 & \simeq &
  -\frac{3\eta}{\tilde{c}^2}
  - \frac{1}{\tilde{c}}
  \left[\frac{V''\phi}{V'}+(\tilde{c}-2)\right]
  = -\frac{3\eta_c}{(c+2)^2},
\end{eqnarray}
%======================================%
where we have set $\tilde{c}=c+2$ in the last equality. Therefore,
$|[\dot{\pi}-(\tilde{c}-2)H\pi]/(\tilde{c}H\pi)|\ll 1$ is equivalent to 
(\ref{eqn:tilde-eta_c}), and the second approximate equation is
justified.

%======================================%
% Angular Fluctuations
%======================================%
\section*{Appendix B: Angular Fluctuations of Mobile Brane  in the Warped Geometry}

First we recall the KS geometry. The Klebanov-Strassler geometry has the
simple ansatz: 
%============< EQUATION >==============%
%
\begin{equation}
 ds^2 = h^{-1/2}(\tau)\eta_{\mu\nu}dx^{\mu}dx^{\nu}+ h^{1/2}(\tau)ds_6^2,
\end{equation}
%======================================%
where $x^{\mu}$ ($\mu=0,\cdots,3$) are $4$-dimensional coordinates and 
$ds_6^2$ is the metric of the deformed
conifold
%============< EQUATION >==============%
%
\begin{equation}
 ds_6^2 = \frac{\epsilon^{4/3}}{2}K(\tau)
  \left[ \frac{1}{3K^3(\tau)}\left(d\tau^2+(g^5)^2\right)
   + \cosh^2\left(\frac{\tau}{2}\right)\left((g^3)^2+(g^4)^2\right)
   + \sinh^2\left(\frac{\tau}{2}\right)\left((g^1)^2+(g^2)^2\right)
  \right]. \label{eqn:metric-deformed-conifold}
\end{equation}
%======================================%
Here, 
%============< EQUATION >==============%
%
\begin{equation}
 K(\tau)= \frac{(\sinh(2\tau)-2\tau)^{1/3}}{2^{1/3}\sinh\tau},
\end{equation}
%======================================%
and $g^i$ ($i=1,\cdots,5$) are orthonormal basis defined by 
%============< EQUATION >==============%
%
\begin{eqnarray}
 g^1 & = & \frac{e^1-e^3}{\sqrt 2},\quad
  g^2 = \frac{e^2-e^4}{\sqrt 2}, \nonumber \\
 g^3 & = & \frac{e^1+e^3}{\sqrt 2},\quad
  g^4 = \frac{e^2+ e^4}{\sqrt 2},\quad 
  g^5 = e^5,
\end{eqnarray}
%======================================%
where
%============< EQUATION >==============%
%
\begin{eqnarray}
 e^1 & \equiv & - \sin\theta_1 d\phi_1, \quad
  e^2 \equiv d\theta_1, \nonumber \\
 e^3 & \equiv & 
  \cos\psi\sin\theta_2 d\phi_2-\sin\psi d\theta_2, \nonumber\\
 e^4 & \equiv & \sin\psi\sin\theta_2 d\phi_2+\cos\psi d\theta_2, 
  \nonumber \\
 e^5 & \equiv & d\psi + \cos\theta_1 d\phi_1+ \cos\theta_2 d\phi_2.
\end{eqnarray}
%======================================%
Because of the warp factor $h^{-1/2}(\tau)$, this geometry is often
called the warped deformed conifold. The R-R $3$-form field strength
$F_3$ and the NS-NS $2$-form potential $B_2$ also have the $Z_2$
symmetric (($\theta_1$, $\phi_1$) $\leftrightarrow$ ($\theta_2$,
$\phi_2$)) ansatz: 
%============< EQUATION >==============%
%
\begin{eqnarray}
 F_3 & = & \frac{M\alpha'}{2} \left\{g^5\wedge g^3\wedge g^4 + d [ F(\tau)
	 (g^1\wedge g^3 + g^2\wedge g^4)]\right\}, \nonumber\\
 B_2 & = & \frac{g_s M\alpha'}{2}
  [f(\tau) g^1\wedge g^2 +  k(\tau) g^3\wedge g^4 ],
\end{eqnarray}
%======================================%
where $F(0)=0$ and $F(\infty)=1/2$. For this ansatz with the additional
condition 
%============< EQUATION >==============%
%
\begin{equation}
 g_s^2F_3^2 = H_3^2,
\end{equation}
%======================================%
we can consistently set the dilaton $\phi$ and the R-R scalar $C_0$
to zero. The BPS saturated solution found by Klebanov and
Strassler is 
%============< EQUATION >==============%
%
\begin{eqnarray}
 F(\tau) & = & \frac{\sinh\tau -\tau}{2\sinh\tau}, \nonumber\\
 f(\tau) & = & \frac{\tau\coth\tau-1}{2\sinh\tau}(\cosh\tau-1), \nonumber\\
 k(\tau) & = & \frac{\tau\coth\tau-1}{2\sinh\tau}(\cosh\tau+1),
\end{eqnarray}
%======================================%
and 
%============< EQUATION >==============%
%
\begin{equation}
 h(\tau) = 2^{2/3}\cdot (g_sM\alpha')^2\epsilon^{-8/3}I(\tau), 
\end{equation}
%======================================%
where 
%============< EQUATION >==============%
%
\begin{equation}
 I(\tau) = 
  \int_\tau^\infty dx \frac{x\coth x-1}{\sinh^2 x}
  (\sinh(2x)-2x)^{1/3}.
\end{equation}
%======================================%
For this solution, 
%============< EQUATION >==============%
%
\begin{equation}
 C_4 = h^{-1}dx^0\wedge dx^1\wedge dx^2\wedge dx^3
\end{equation}
%======================================%
in a particular gauge. For large $g_sM$ the curvature is small
everywhere and we can trust the supergravity description.

When $g_sM$ is sufficiently large, we can treat a $D3$-brane as a
probe brane. The action for the probe $D3$-brane is 
%============< EQUATION >==============%
%
\begin{equation}
 S_{D3} = -T_3\int d^4\xi e^{-\phi}
  \sqrt{-\det(G_{\alpha\beta}-B_{\alpha\beta})}
  + T_3\int d^4\xi C_4,
\end{equation}
%======================================%
where $\xi^{\alpha}$ ($\alpha=0,\cdots,3$) are intrinsic coordinates on
the $D3$-brane, $T_3$ is the tension and 
%============< EQUATION >==============%
%
\begin{equation}
 G_{\alpha\beta} = G_{MN}
  \frac{\partial x^M}{\partial\xi^{\alpha}}
  \frac{\partial x^N}{\partial\xi^{\beta}}, \quad
 B_{\alpha\beta} = (B_{2})_{MN}
  \frac{\partial x^M}{\partial\xi^{\alpha}}
  \frac{\partial x^N}{\partial\xi^{\beta}}.
\end{equation}
%======================================%
In the following we shall adopt a gauge in which brane coordinates
$\xi^{\alpha}$ coincide with $x^{\alpha}$:
%============< EQUATION >==============%
%
\begin{equation}
 x^{\alpha} = \xi^{\alpha},\quad \psi^m = \psi^m(\xi^{\alpha}), 
\end{equation}
%======================================%
where $\{\psi^m\}$ ($m=5,\cdots,10$) represents
$\{\tau,\psi,\theta_1,\phi_1,\theta_2,\phi_2\}$. In the
non-relativistic limit, 
%============< EQUATION >==============%
%
\begin{equation}
 S_{D3} = -\frac{T_3}{2}\int d^4\xi 
  \gamma_{mn}\eta^{\alpha\beta}
   \frac{\partial\psi^m}{\partial\xi^{\alpha}}
   \frac{\partial\psi^n}{\partial\xi^{\beta}}. 
\end{equation}
%======================================%
where $\gamma_{mn}d\psi^md\psi^n=ds_6^2$ is the metric of the deformed
conifold given by (\ref{eqn:metric-deformed-conifold}). 
For a cosmological background, if the energy scale is sufficiently low,
we can replace $\eta_{\mu\nu}$ by $g_{\mu\nu}$ to obtain
%============< EQUATION >==============%
%
\begin{equation}
 S_{D3} = -\frac{T_3}{2}\int d^4\xi \sqrt{-g}
  \gamma_{mn}g^{\mu\nu}
   \frac{\partial\psi^m}{\partial x^{\mu}}
   \frac{\partial\psi^n}{\partial x^{\nu}}. 
\end{equation}
%======================================%

For large $\tau$, $K(\tau)$ and $I(\tau)$ behaves as
%============< EQUATION >==============%
%
\begin{equation}
 K(\tau) \simeq 2^{1/3}e^{-\tau/3}, \quad
  I(\tau) \simeq \frac{3}{2^{1/3}}e^{-4\tau/3}.
\end{equation}
%======================================%
Thus, we have 
%============< EQUATION >==============%
%
\begin{equation}
 \gamma_{mn}\partial\psi^m\partial\psi^n
  = \frac{\epsilon^{4/3}}{2^{2/3}}e^{2\tau/3}
  \left[\frac{1}{6}\left(d\tau^2+(g^5)^2\right)
   +(g^3)^2+(g^4)^2+(g^1)^2+(g^2)^2 \right].
\end{equation}
%======================================%
By introducing the canonically normalized variable $\phi$ as
%============< EQUATION >==============%
%
\begin{equation}
 \phi = \frac{\epsilon^{2/3}\sqrt{3T_3}}{2^{5/6}}e^{\tau/3}, 
\end{equation}
%======================================%
we obtain
%============< EQUATION >==============%
%
\begin{equation}
 T_3\gamma_{mn}\partial\psi^m\partial\psi^n
  = \partial\phi\partial\phi
  + \frac{2}{3}\phi^2
  \left[\frac{1}{6}(g^5)^2 +(g^3)^2+(g^4)^2+(g^1)^2+(g^2)^2 \right]. 
\end{equation}
%======================================%
Therefore, 
%============< EQUATION >==============%
%
\begin{eqnarray}\label{full}
 S_{D3} & = & -\frac{1}{2}\int d^4\xi \sqrt{-g}
  g^{\mu\nu}\left\{
	     \partial_{\mu}\phi\partial_{\nu}\phi 
+ \frac{2}{3}\phi^2
  \left[\frac{1}{6}g^5(\partial_{\mu})g^5(\partial_{\nu})
    \right.\right.  \nonumber\\
 & & \left. \left.
   +g^3(\partial_{\mu})g^3(\partial_{\nu})
   +g^4(\partial_{\mu})g^4(\partial_{\nu})
   +g^1(\partial_{\mu})g^1(\partial_{\nu})
   +g^2(\partial_{\mu})g^2(\partial_{\nu})
   \right]  \right\}.
\end{eqnarray}
%======================================%

Note that the kinetic term of angular coordinates has the overall factor
$\phi^2$. In more general situation where the KS geometry is modified,
the overall factor will be the angular component of the $6$-dimensional
metric multiplied by the $D3$-brane tension $T_3$.

For simplicity, we continue our analysis with single angular coordinate.
Result will be similar for the full geometry (\ref{full}).
Let us consider a scalar field $\theta$ coupled to the conformal
inflaton $\phi$ described by 
%============< EQUATION >==============%
%
\begin{equation}
 I = \int d^4x\sqrt{-g}
  \left[ \frac{M_p^2R}{2}
   -\frac{1}{2}\partial^{\mu}\phi\partial_{\mu}\phi
   - \frac{1}{2}\phi^2\partial^{\mu}\theta\partial_{\mu}\theta 
   -V(\phi) - \frac{\xi}{2}R\phi^2\right],
\end{equation}
%======================================%
where $\xi=1/6$ and the potential $V(\phi)$ is assumed to satisfy the
conditions for conformal inflation. During the conformal inflation, 
%============< EQUATION >==============%
%
\begin{equation}
 \phi \simeq \frac{\varphi_0}{a},
\end{equation}
%======================================%
where $\varphi_0$ is a constant. Hence, the equation of motion for
$\theta$ is
%============< EQUATION >==============%
%
\begin{equation}
 \ddot{\theta}_{\vec{k}} +  H\dot{\theta}_{\vec{k}} 
  + \frac{\vec{k}^2}{a^2}\theta_{\vec{k}} = 0 \ .
\end{equation}
%======================================%
Switching to the conformal time brings this equation to
the wave equation in the flat spacetime.
\begin{equation}
 {\theta}''_{\vec{k}} 
  + {\vec{k}^2}\theta_{\vec{k}} = 0 \ .
\end{equation}
Thus, angular fluctuations are effectively described by the
equation for the  conformal fields and, thus,
cannot be responsible for the cosmological modulated fluctuations.

%======================================%
%<<<<<<<<<<<< REFERENCES >>>>>>>>>>>>>>%
%======================================%

\end{document}